\documentclass[pdftex,aps,prb,twocolumn,showpacs,letterpaper]{revtex4}
\usepackage{graphicx,rotating,epsfig,amsmath,color}
\usepackage[bookmarks=false]{hyperref}
\pdfoutput=1
\newcommand{\kv}{\mathbf{k}} \newcommand{\qv}{\mathbf{q}}
\newcommand{\Qv}{\mathbf{Q}} \newcommand{\be}{\begin{equation}}
\newcommand{\ee}{\end{equation}} \newcommand{\bea}{\begin{eqnarray}}
\newcommand{\eea}{\end{eqnarray}} 
 \newcommand{\bwt}{\begin{widetext}}
\newcommand{\ewt}{\end{widetext}} \newcommand{\ham}{\mathcal{H}}
\newcommand{\fc}{\mathcal{F}} \newcommand{\ra}{\rangle}
\newcommand{\la}{\langle} \newcommand{\bsb}{\begin{subarray}}
\newcommand{\esb}{\end{subarray}}

\begin{document}
\title{Holstein polaron: the effect of multiple phonon modes}

\author{Lucian Covaci and Mona Berciu} 

\affiliation{ Department of Physics and Astronomy, University of
  British Columbia, Vancouver, BC, Canada, V6T~1Z1}

\begin{abstract}
We generalize the Momentum Average approximations MA$^{(0)}$ and
MA$^{(1)}$ to study the effects of coupling to multiple optical phonons
on the properties of a Holstein polaron. As for a single phonon
mode, these approximations are numerically very efficient. They become
exact for very weak or very strong couplings, and are highly accurate
in the intermediate regimes, {\em e.g.} the spectral weights
obey exactly the first six, respectively eight, sum rules. Our results
show that the effect on ground-state properties is cumulative in
nature. In particular, if the effective coupling to one mode is much
larger than to the others, this mode effectively determines the GS
properties. However, even very weak coupling to a second phonon mode has
important non-perturbational effects on the higher energy
spectrum, in particular on the dispersion and the phonon statistics of the
polaron band.
\end{abstract}

 \date{\today} \maketitle

The coupling of electrons to phonons is a widely studied problem,
because it leads to many interesting phenomena such as conventional
superconductivity or the formation of polarons (composite objects
comprised of an electron and the surrounding phonon cloud), important
in several classes of materials. As a recent example, results from
angle-resolved photoemission spectroscopy\cite{arp} (ARPES) have lead
to new discussions about possible polaronic effects in
high-temperature superconductors.\cite{kyle}

Most theoretical studies of polaron properties are of the Holstein
model with a single optical phonon mode,\cite{alexandrov, feshke} even
though complex materials have many optical and acoustic phonons. The
reason is that usually there is one optical mode to which the coupling
is strongest, and one assumes that the effects of the other modes are
perturbationally small. Also, the efficiency of various numerical
methods\cite{feshke} such as exact diagonalization, diagrammatic Monte
Carlo and variational methods employed for obtaining results in the
intermediate coupling regime, where no exact solutions are known,
suffers when the Hilbert space is enlarged by addition of multiple
phonon modes.

Recently, the so-called Momentum Average (MA) analytical
approximation\cite{berciu:prl,goodvin:2006} has been shown to be
highly accurate over most of the parameter space of the Holstein
polaron problem, while requiring a numerically trivial effort irrespective of
the dimensionality of the problem or the strength of the
coupling. Moreover, its accuracy can be systematically
improved.\cite{berciu:2007} Such fast but accurate methods are useful
for a quick survey of polaron properties in various regimes, which can
then be followed by quantitatively more accurate, but significantly
more time and resource consuming numerical simulations.

In this Rapid Communication we show how these approximations can be
generalized to deal with multiple phonon modes, without loss of
accuracy when compared to the single-mode results. This allows us to
study easily the effects of multiple phonon modes on the polaron
properties. As we show below, while for ground-state properties these
effects are rather trivial, the higher-energy spectrum is significantly
modified by additional phonon modes, even if coupling to them is
perturbationally small.

The generalized Holstein Hamiltonian\cite{holstein} of interest is:
\begin{eqnarray}
\nonumber \ham=&&\sum_\kv \epsilon_\kv c_\kv^\dagger c_\kv +
\sum_{\qv,\alpha} \Omega_\alpha b_\qv^{\alpha \dagger} b_\qv^\alpha \\
&&+\sum_{\alpha,\kv,\qv} \frac{g_\alpha}{\sqrt{N}} c_{\kv-\qv}^\dagger
c_{\kv} (b^{\alpha \dagger}_\qv + b^{\alpha }_{-\qv}).
\label{hamil} 
\end{eqnarray}
The first term describes a free electron on a $d$-dimensional lattice
with $N$ sites (its spin is irrelevant, thus the spin index is dropped),
the second describes the optical phonon modes, and the third describes
the electron-phonons couplings. Momenta sums are over the Brillouin
zone.

We focus on finding the polaron's Green's
function:\cite{berciu:prl,goodvin:2006} \be G(\kv,\omega)=\la 0| c_\kv
\hat{G}(\omega) c^\dagger_\kv |0\ra, \ee where
$\hat{G}(\omega)=[\omega-\ham+i\eta]^{-1}$ is the usual resolvent
($\hbar =1$) and $|0\ra$ is the vacuum. The poles of this Green's
function mark the polaron spectrum, and ground-state (GS) energies,
effective masses, quasiparticle ($qp$) weights and average phonon
numbers can then be calculated as discussed in
Ref. \onlinecite{goodvin:2006}. The spectral weight $A(\kv, \omega) =
- {1\over \pi} G(\kv,\omega)$ can also be directly compared against
ARPES results.

To simplify notation, we first assume that there are only two
phonon modes, and rename their operators as $b_\qv$ (mode 1) and
$B_\Qv$ (mode 2). The generalization to more phonon modes is discussed below. 
We use repeatedly Dyson's identity
$\hat{G}(\omega)=\hat{G}_0(\omega)+\hat{G}(\omega) \hat{V}
\hat{G}_0(\omega)$, where
$\hat{G}_0(\omega)=[\omega-\ham_0+i\eta]^{-1}$ corresponds to the
non-interacting Hamiltonian, to generate an infinite system of coupled
equations involving $G(\kv,\omega)$ and the generalized Green's
functions $F_{nm}(\kv,\qv_1,\dots,\qv_n;\Qv_1,\dots,\Qv_m;\omega)=\la 0| c_\kv
\hat{G}(\omega) c_{\kv_T}^\dagger b^{\dagger}_{\qv_1}
\dots b^{\dagger}_{\qv_n}B^{\dagger}_{\Qv_1}
\dots B^{\dagger}_{\Qv_m}|0 \ra$. Here $\kv_T = \kv-\qv_T-\Qv_T$ and $
\qv_T=\sum_{i=1}^{n}\qv_i$, $\Qv_T=\sum_{j=1}^{m}\Qv_j$. Arguments
identical to those of Ref. \onlinecite{berciu:2007} show that all
functions $F_{nm}$ are proportional to $G(\kv,
\omega)$. It is thus more convenient to work with  the rescaled functions
$f_{nm}(\kv,\{\qv\},\{\Qv\},\omega)=N^{(n+m)/2} 
F_{nm}(\kv,\{\qv\},\{\Qv\},\omega)/G(\kv,\omega)$, where we use the
shorthand notation 
$\{\qv\}\equiv {\qv_1},
\dots,{\qv_n}$, etc.

The solution has the standard form
\be G(\kv,
\omega)= \left[\omega - \epsilon_{\kv} -\Sigma(\kv, \omega) + i
\eta\right]^{-1}
\ee  where the exact self-energy is given by
\be
\Sigma(\kv,\omega)=\frac{g_1}{N}\sum_{\qv_1}f_{10}(\kv,\qv_1,\omega) +
\frac{g_2}{N}\sum_{\Qv_1}f_{01}(\kv,\Qv_1,\omega)
\label{selfe_1}
\ee
In terms of the new sets $\{\qv\}_i \equiv {\qv_1},
\dots,{\qv_{i-1}},
{\qv_{i+1}},\dots,{\qv_n}$ and $\{\qv\}_{n+1} \equiv {\qv_1},
\dots,{\qv_n}, {\qv_{n+1}}$,  the functions $f_{nm}$ are the
solutions of  the following recurrence relations:
\begin{widetext}
\begin{eqnarray}
f_{nm}(\{\qv\},\{\Qv\})=& &
G_0(\kv_T,\omega-n\Omega_1-m\Omega_2)  
\left[ g_1 \sum_{i=1}^n f_{n-1,m}(\{\qv\}_i,\{\Qv\})+ g_2
\sum_{j=1}^m f_{n,m-1}(\{\qv\},\{\Qv\}_j) \right. \nonumber \\
& & \label{rec} \left. +
\frac{g_1}{N}\sum_{\qv_{n+1}}f_{n+1,m}(\{\qv\}_{n+1},\{\Qv\}) +
\frac{g_2}{N}\sum_{\Qv_{m+1}}f_{n,m+1}(\{\qv\},\{\Qv\}_{m+1})\right]
\end{eqnarray}
\end{widetext}
where the dependence on $\kv, \omega$ is implicitly assumed for all $f_{nm}$, 
$G_0(\kv, \omega)= (\omega - \epsilon_{\kv} + i \eta)^{-1}$ is the
free propagator, and $f_{00}\equiv 1$ by definition. These are the
generalization of the equivalent single-mode equations of
Ref. \onlinecite{berciu:2007}. 

As for the single-mode problem, the MA$^{(0)}$ approximation is
obtained by replacing in the {\em r.h.s} of 
Eqs. (\ref{rec})
$$ 
G_0(\kv_T,\omega-n\Omega_1-m\Omega_2) \rightarrow
\bar{g}_0(\omega-n\Omega_1-m\Omega_2)\equiv \bar{g}_{nm}(\omega)
$$
where the momenta averages
$$
\bar{g}_0(\omega) = {1\over N} \sum_{\kv}^{} G_0(\kv,\omega)
$$
are simple known functions.\cite{note} In terms of the functions
\be
\fc_{nm}(\omega)=\frac{1}{N^{m+n}}\sum_{\{\qv\},\{\Qv\}}
f_{nm}(\kv,\{\qv\},\{\Qv\},\omega),  \ee
the momentum-independent MA$^{(0)}$ self-energy is:
$$\Sigma_{MA^{(0)}}(\omega)=g_1\fc_{10}(\omega)+g_2\fc_{01}(\omega),$$ 
while the recurrence relations (\ref{rec}) take the simpler form
$\fc_{nm}(\omega)=\bar{g}_{nm}(\omega)[ng_1\fc_{n-1,m}(\omega)+ 
mg_2\fc_{n,m-1}(\omega) +g_1\fc_{n+1,m}(\omega) +
g_2\fc_{n,m+1}(\omega)]$ (of course,
$\fc_{00}=1$). 

Such recursive equations were previously solved in a different context
by Cini {\em et al.},\cite{cini} but their solution cannot be
generalized to more than two phonon modes, nor to MA$^{(1)}$ or higher
levels (see below). We have found an alternative solution without
these shortcomings. First, we rewrite these recurrence relations in
matrix form: \be V_k=A_kV_{k-1}+B_kV_{k+1},
\label{recvec}
\ee where the vector $V_k=(\fc_{k,0};
\fc_{k-1,1};\ldots;\fc_{1,k-1};\fc_{0,k})^T$ contains all $k+1$
functions corresponding to a total of $k$ phonons. $A_k$ is a matrix
of size $k+1 \times k$ with the only non-zero elements
$(A_k)_{i,i}=(k-i)g_1\bar{g}_{k-i,i}(\omega)$ and
$(A_k)_{i+1,i}=(i+1)g_2\bar{g}_{k-1-i,i+1}(\omega)$, $\forall i=0,
k-1$. Similarly, $B_k$ is a matrix of size $k+1 \times k+2$ with the
only non-zero elements $(B_k)_{i,i}=g_1\bar{g}_{k-i,i}(\omega)$ and
$(B_k)_{i,i+1}=g_2\bar{g}_{k-i,i}(\omega)$, $\forall i=0, k$.
Dependence on $\omega$ is again implicitly assumed everywhere.

The solution is $V_k = M_k V_{k-1}$ with $V_0=(1)$, where \be
 M_k=\cfrac{1}{ 1-B_k \cfrac{1}{ 1- B_{k+1}
 \cfrac{1}{1-\ldots}A_{k+2}}A_{k+1} }A_k.
\label{cfracma0}
\ee is a continued fraction of matrices of increasing size.  The self
energy is $\Sigma_{MA^{(0)}}(\omega) = (g_1,g_2)V_1 =
(g_1,g_2)M_1$. The continued fractions become convergent if truncated
at levels $N \approx g_1^2/\Omega_1^2+g_2^2/\Omega_2^2$,
i.e. when one keeps contributions from $f_{nm}$ corresponding to
expected average numbers of phonons in the cloud. The generalization
to more phonon modes is straightforward.  $V_k$ again contains all
Green's functions with a fixed number of phonons, and the interaction
links it only to $V_{k\pm1}$.  The matrices $A_k$ and $B_k$ have
non-vanishing elements only on a number of diagonals equal to the
number of modes. The dimension of these matrices increases now faster
with increasing $k$, but it is still much less severe than the
corresponding increase in numerical simulations. Also, note that the
MA calculation is equally simple in any dimension, the only change
being in the expression used for $\bar{g}_0(\omega)$.\cite{note}

The analysis of the diagrammatic and variational meaning of
MA$^{(0)}$ and the sum rules it obeys, is identical to that for
the single-mode case,\cite{berciu:prl,goodvin:2006,berciu:2007} and we
do not repeat it. It again proves its accuracy over the
entire parameter space, as long as $\Omega_i/t >0.1, \forall
i$. However, MA$^{(0)}$ fails to correctly predict the polaron+one
phonon continuum.\cite{goodvin:2006,reps} To remedy this, we  use
MA$^{(1)}$ or a higher level approximation.\cite{berciu:2007} In
MA$^{(1)}$,  Eqs.~(\ref{rec}) for $f_{01}$ and $f_{10}$ are left
unchanged, and the momentum average is made only for $f_{nm}$ with
$n+m\ge 2$. As shown in Ref. \onlinecite{berciu:2007}, MA$^{(1)}$
correctly predicts the polaron+one phonon
continuum, besides giving small improvements in the accuracy of
various other quantities, {\em e.g.} the number of exactly satisfied
sum rules increases from six to eight. The derivation of
$\Sigma_{MA^{(1)}}(\omega)$ follows identically that in
Ref. \onlinecite{berciu:2007}, with the only difference that continued
fractions there correspond to continued fractions of matrices [like in
Eq. (\ref{cfracma0})]
here. We will present the details of this straightforward
derivation elsewhere,\cite{covaci} instead focusing here on results.

\begin{figure}[t]
\includegraphics[width=\columnwidth]{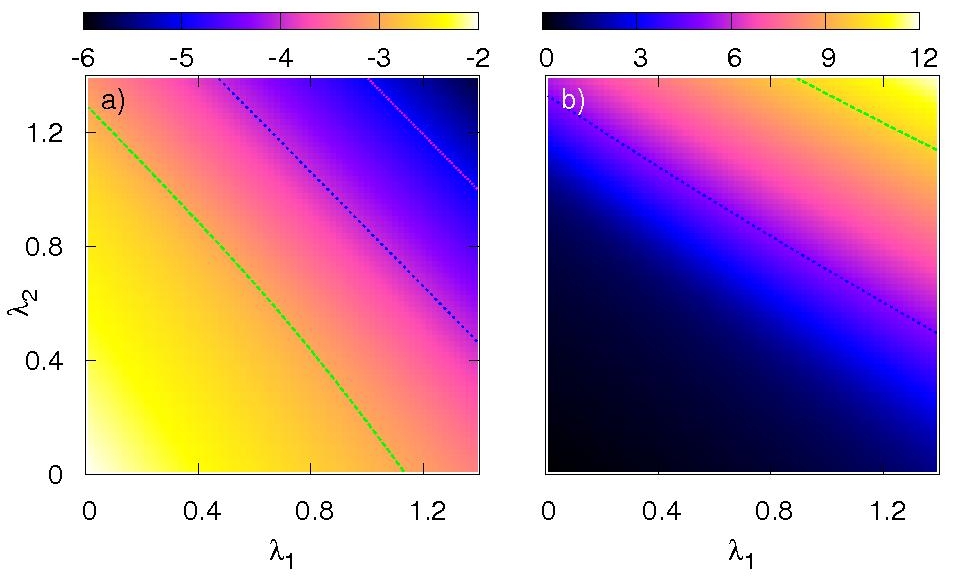}
 \caption{(color online) (a) MA$^{(1)}$ GS  energy and (b) $\ln(m^*/m)$,
 where $m^*$ is the polaron effective mass, {\em vs.}  coupling constants
 $\lambda_1$ and $\lambda_2$, for  $\Omega_1=0.7t$,  $\Omega_2=0.3t$ and $d=1$.}
\label{fig:1}
\end{figure}

All results shown are for two phonon modes and $d=1$, which suffices
to uncover the essential new physics. Results for higher $d$ and more
phonon modes will be presented elsewhere.\cite{covaci} We begin by discussing
GS properties such as the energy $E_{GS}$ and effective mass $m^*$,
shown in Fig. \ref{fig:1} as functions of the effective couplings
$\lambda_i=g_i^2/(2dt\Omega_i)$, $i=1,2$. Note that the $qp$ weight is
$Z_0=m/m^*$,  where $m$ is the bare electron
mass.\cite{goodvin:2006} The ``equipotential'' lines drawn show that
$E_{GS}$ is well described as a function of only $\lambda_{\rm
eff}=\sum_{i}^{}\lambda_i$, whereas $m_*$ is a function of
$\sum_{i}^{}\lambda_i/\Omega_i$. If $\Omega_i=\Omega$ for all modes, these
can be proved to be {\em exact} results.\cite{covaci} In the strong
coupling limit one also expects $E_{GS} = -\sum_{i}^{}g_i^2/
\Omega_i$, $\ln Z_0 \propto  -\sum_{i}^{}\lambda_i/\Omega_i$,
supporting the same conclusion. To a good extent the only
effect of having $\Omega_1 \ne \Omega_2$ is to change the slope of the $m^*$
``equipotentials'' from the $45^o$ found when $\Omega_1=\Omega_2$,
although some slight deviations from linearity are also seen in
$E_{GS}$ when
either $\lambda_i \ll 1$.

We conclude that GS properties can be well
understood in cumulative terms, for instance the energy is  that of a
polaron coupled to a single phonon with 
$\lambda_{\rm eff}$. The crossover from large to small-polaron
behavior is therefore expected when $\lambda_{\rm eff}\approx
1$.\cite{alexandrov, feshke} As a result,  it is possible to have
small-polaron behavior even if each individual phonon mode is weakly
coupled to the electron ($\lambda_i < 1$). However, in cases where one
mode (say, mode 1) is indeed much more strongly coupled than all
others, $\lambda_1 \gg \lambda_i$, $i=2,\dots$, then $\lambda_{eff}
\approx \lambda_1$ and one can, to a good extent, ignore the small
cumulative effect from the other modes. 

\begin{figure}[t]
\includegraphics[width=\columnwidth]{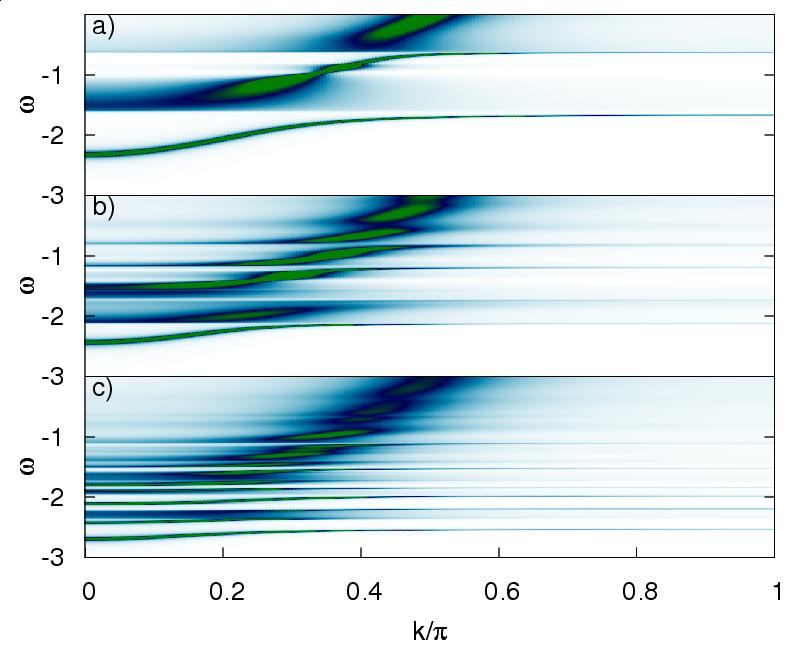}
 \caption{(color online) $A(k,\omega)$ from MA$^{(1)}$ for $d=1$, and
 for $\Omega_1=0.7t$, $\Omega_2=0.3t$, 
 $\lambda_1=0.4$ and a) $\lambda_2=0.0$, b) $\lambda_2=0.2$, c)
 $\lambda_2=0.6$.}
\label{fig:2}
\end{figure}

This conclusion, however, does not generally hold for higher-energy properties,
as we show now. In Fig.~\ref{fig:2} we plot the $d=1$ spectral
function $A(k,\omega)$ {\em vs.} $k$ and $\omega$. The effective
coupling to the first mode, of frequency $\Omega_1/t=0.7$, is kept
constant to a fairly low value $\lambda_1=0.4$. In panel (a),
coupling to the second mode, of energy $\Omega_2/t=0.3$, is zero, so
this is effectively a one-phonon problem. As expected, at low energies
we see the polaron band, of width $\Omega_1$, followed above
$E_{GS}+\Omega_1$  by the polaron+one-phonon continuum, and other
features at even higher energies.\cite{alexandrov, feshke,
  berciu:2007} 

Addition of even very weak coupling to a second phonon mode changes
things considerably, as shown in panel (b) for $\lambda_2 = 0.2$. If
$\Omega_2 < \Omega_1$ (as chosen here), the polaron band width is
changed to $\Omega_2$, even though the GS energy and effective mass
are not much affected (see previous discussion). This significant
change is not so surprising if one considers the origin of the
polaron+one phonon continuum: it corresponds to states where one
phonon is created far from the polaron. As a result, they interact
little and the total energy is just the sum of the two. If there are
several phonon modes, the continuum will be defined by the mode with
the lowest frequency $\Omega_{\rm min}$, irrespective of whether this is the
mode most strongly coupled to the electron or not. Because of this,
the polaron band cannot be wider than $\Omega_{\rm min}$. 

\begin{figure}[t]
\centering
\includegraphics[width=\columnwidth]{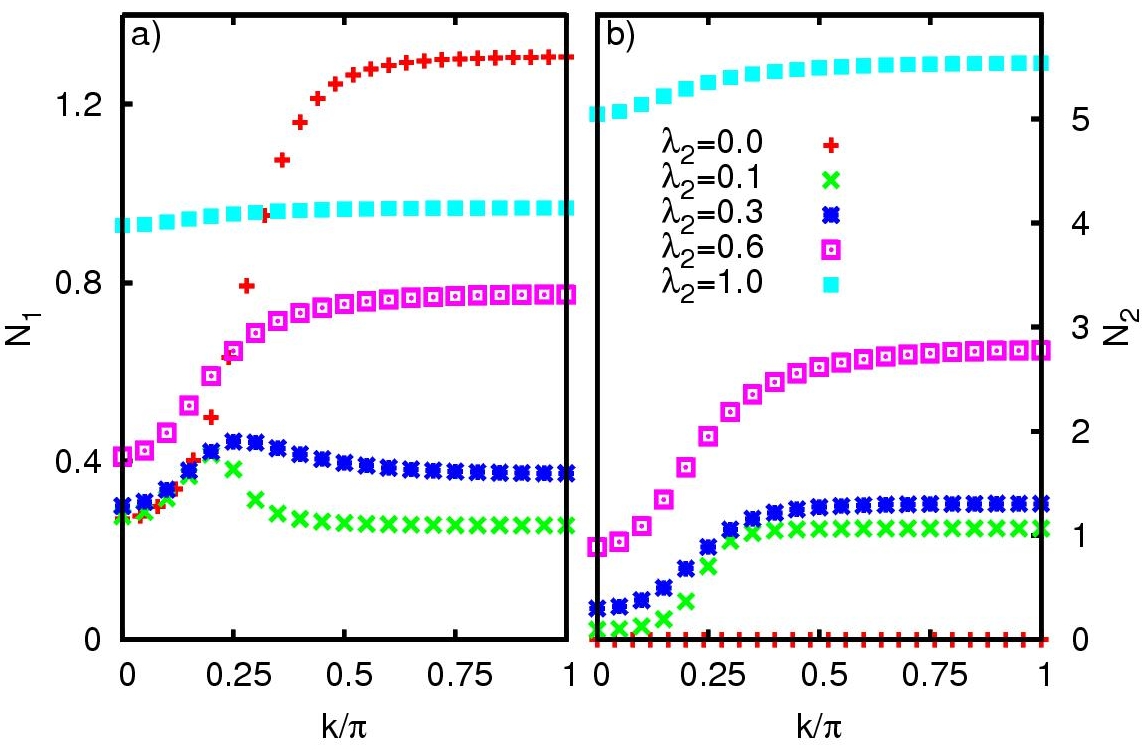}
\caption{(color online) Number of phonons of either type in the polaron cloud {\em vs.}
$\lambda_2$, for $\Omega_1=0.7t$, $\Omega_2=0.3t$ and $\lambda_1=0.4$.}
\label{fig:4}
\end{figure}

This interpretation is confirmed by phonon statistics, shown in
Fig.~\ref{fig:4}. Here we plot average numbers $N_1$ and $N_2$ of
phonons of either type in the polaron cloud, as a function of the
polaron momentum $k$. These are calculated using the Hellman-Feynman
theorem.\cite{feynman,goodvin:2006} Again, the coupling to the first mode is
kept constant at $\lambda_1=0.4$. If $\lambda_2=0$ ( $+$ symbols), we
see that $N_1$ increases from a small value at $k=0$ to just above $1$
for  $k>\pi/2$. This shows, as expected,  that while  around $k=0$ the large-polaron
is essentially similar to a free electron, for $k> \pi/2$ the largest
contribution to the polaron comes from electron+one phonon states. Of
course, $N_2=0$  in this case. As $\lambda_2$ is turned on but is still
small ($\lambda_2=0.1, 0.3$, $\times$ symbols) there is little change near $k=0$,
however the changes at higher momenta are dramatic: $N_1$
decreases by 1 whereas $N_2$ increases by 1 (note the different
scales). This confirms that it is now the second type of phonon that controls
the nature of the polaron at large $k$ values, even though $\lambda_2 <
\lambda_1$. Once $\lambda_2 > \lambda_1$, the second phonon completely
dominates behavior. For $\lambda_2 = 0.6$ (square symbols) we have
$\lambda_{\rm eff} = \lambda_1 + \lambda_2 =1$, and the polaron is in
the crossover regime, whereas for $\lambda_2 = 1.0, \lambda_{\rm eff}=1.4$ the polaron is
firmly in the small-polaron regime. In the latter case, we expect to
see less and less $k$ dependence, since the polaron cloud becomes limited to the
site on which the electron resides. This behavior is indeed observed for $N_2$,
but also for $N_1$. This is because once the
polaron is localized at a site (because of strong coupling to mode 2)
it will automatically also shift the equilibrium position for mode 1
phonons at that site, resulting in the creation of a finite number of
such bare phonons.

\begin{figure}[t]
\centering \includegraphics[width=\columnwidth]{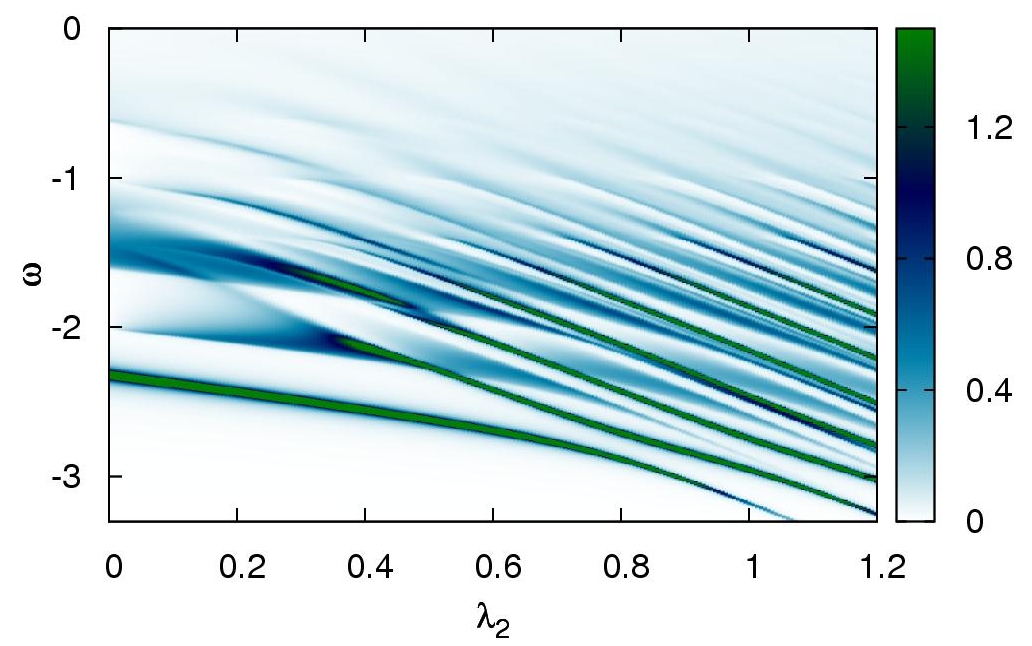}
\caption{(color online) $A(k=0,\omega)$ {\em vs.} $\omega$ and $\lambda_2$, for
$\Omega_1=0.7t$, $\Omega_2=0.3t$ and $\lambda_1=0.4t$.}
\label{fig:5}
\end{figure}

If there is only one phonon mode coupled to the electron, one can use
ARPES data to extract its frequency $\Omega$ from the location of the
``discontinuity'' in the dispersion, at weak coupling, or from average
distance between eigenstates at strong couplings, where the
Lang-Firsov spectrum appears.\cite{firsov} The effective mass
determines $\lambda$, so $g$ can also be extracted. Our results show
that this procedure is usually wrong in the case of multiple phonon
modes, even if one expects coupling to one of them to be dominant. It
only works if this particular mode also happens to have the lowest
frequency, else one will underestimate its $\Omega$ and overestimate
$g$. This point is clearly demonstrated in Fig.~\ref{fig:5}, which
shows that the $\Omega_{\rm min}$ phonon defines the location of the
low-energy (not GS) features, irrespective of its coupling. Of course,
one may hope to see the continua due to the other phonon modes (see
also Fig.~\ref{fig:2}.c) and thus be able to identify their
frequencies. This is probably unlikely, due to broadening in real data
(note that temperature dependence would also be determined by the
$\Omega_{\rm min}$ phonon, not by the dominant one). Even if the
$\Omega_i$ are identified, finding all $g_i$ is generally impossible,
unless we know that one dominates, and we know which one that is.  The
only simple case is if {\em all} phonons have roughly equal
frequencies, when one can treat them as a single mode with coupling
$g^2_{\rm eff} = \sum_{i}^{}g_i^2$.

To summarize, we have found a generalization of the simple, yet
accurate Momentum Average approximations to the problem of
Holstein-type coupling to multiple phonon modes. Our results show that
even perturbationally weak coupling to a second phonon can lead to
essential changes of the spectral weight, if its frequency is less
than that of the dominant phonon. In such cases, the simple
way of extracting the electron-phonon coupling from ARPES data is
likely to lead to wrong values.

{\em Acknowledgments}: We thank G. A. Sawatzky
and F. Marsiglio for useful discussions. This work was supported by the A. P. Sloan
Foundation, CIfAR, NSERC and CFI.

\end{document}